\documentclass[pre,aps,12pt]{revtex4}
\usepackage{amssymb}
\usepackage{graphicx}
\begin{document}

\title{Magnetic properties of delta- and kagome-like chains with competing interactions}
\author{D.~V.~Dmitriev}
\author{V.~Ya.~Krivnov}
\email{krivnov@deom.chph.ras.ru}
\affiliation{Institute of
Biochemical Physics of RAS, Kosygin str. 4, 119334, Moscow,
Russia.}
\date{}

\begin{abstract}
We study the delta-chain with spin-$1$ on basal sites and
spin-$\frac{1}{2}$ on apical sites. The Heisenberg interaction
between neighbor basal spins is antiferromagnetic (AF) and the
interaction between basal and apical spins is ferromagnetic (F).
We show that the magnetization curve of this model is the same as
that of the spin-$\frac{1}{2}$ kagome-like chain with competing
Heisenberg interactions. The ground state phase diagram of the
latter as a function of the ratio between the AF and F
interaction, $\alpha $, consists of the ferromagnetic,
ferrimagnetic and singlet phases. We study the magnetic properties
in each ground state phase and analyze the magnetization curves.
We show that there are magnetization plateaus and jumps in
definite regions of value $\alpha$. We compare the magnetic
properties of considered models with those of the
spin-$\frac{1}{2}$ delta chain.
\end{abstract}

\maketitle

\section{Introduction}

\begin{figure}[tbp]
\includegraphics[width=5in,angle=0]{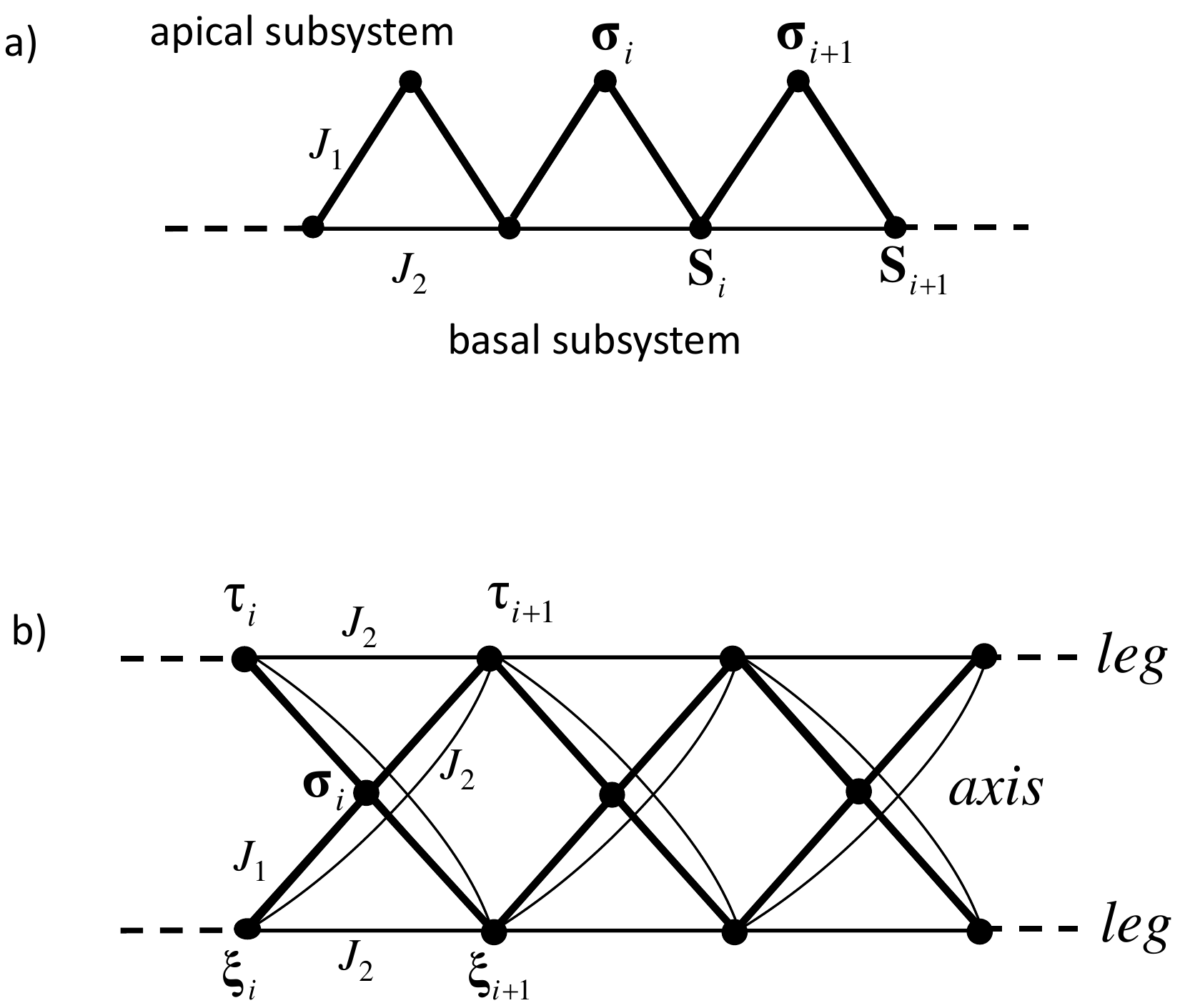}
\caption{Delta-chain (a) and kagome chain (b) spin models.}
\label{Fig_saw}
\end{figure}

Low-dimensional quantum magnets based on a geometrically
frustrated lattice are of considerable interest from both
experimental and theoretical points of view \cite{Diep,Lacrose}.
One of the typical example of these systems is the delta or the
sawtooth chain, i.e. the Heisenberg model on a linear chain of
triangles, as shown in Fig.\ref{Fig_saw}. The Hamiltonian of this
model has the form
\begin{equation}
\hat{H}=J_{1}\sum_{i=1}^{N}(\mathbf{S}_{i-1}+\mathbf{S}_{i})\cdot \mathbf{%
\sigma }_{i}+J_{2}\sum_{i=1}^{N}\mathbf{S}_{i-1}\cdot \mathbf{S}%
_{i}-h\sum_{i=1}^{N}(S_{i}^{z}+\sigma _{i}^{z})  \label{H}
\end{equation}%
where $\mathbf{\sigma}_{i}$ and $\mathbf{S}_{i}$ are the apical
and the basal spin operators with spin quantum numbers $s_{a}$ and
$s_{b}$, correspondingly. The interaction $J_{1}$ acts between
apical and basal spins, while $J_{2}$ is the interaction between
neighbor basal spins. The direct exchange between apical spins is
absent, $h$ is dimensionless magnetic field and $N$ is number of
triangles in the cyclic system.

A quantum delta chain with both antiferromagnetic interactions,
$J_{1}>0$ and $J_{2}>0$ is well studied
\cite{ModernPhysics,Mac,Shulen,Zhit,Zhit2,Zhitomir,Capponi,Balika,Brenig}
and has a number of interesting properties. At the same time, the
delta chain with the ferromagnetic $J_{1}$ and antiferromagnetic
$J_{2}$ interactions (F-AF delta-chain) is very interesting as
well and has attracted interest last time. The $s=\frac{1}{2}$
F-AF delta-chain (with both $s_{a}=$ $s_{b}=\frac{1}{2}$) is a
minimal model describing real compounds, in particular,
malonate-bridged copper complexes
\cite{malonate,malonate1,Tonegawa,Kaburagi} and new kagome
fluoride $Cs_{2}LiTi_{3}F_{12}$ in which the delta-chains host as
magnetic subsystems \cite{Ueda}. Another very interesting example
of the F-AF delta-chain compound is recently synthesized cyclic
complexes $Fe_{10}Gd_{10} $ with $Gd$ and $Fe$ ions as the apical
and basal spins with $s_{a}=\frac{7}{2}$ and $s_{b}=\frac{5}{2}$,
correspondingly \cite{F10}.

Of particular interest is the study of the magnetic properties of
the F-AF model depending on the value of the frustration parameter
$\alpha =\frac{J_{2}}{\left\vert J_{1}\right\vert }$. The ground
state of this model is ferromagnetic for $\alpha
<\frac{s_{a}}{2s_{b}}$. The value $\alpha
_{c}=\frac{s_{a}}{2s_{b}}$ corresponds to the quantum critical
point separating the ground state phases. The properties of the
F-AF delta-chain in the critical point are highly nontrivial. As
it is shown in \cite{DK,KD,ferri,DKRS,Schnack} the ground state
consists of localized magnons and bound magnon complexes which
form macroscopically degenerate ground state manifold. The total
number of the ground states is \cite{DK}:
\begin{equation}
G(N)=2^{N}+2N(s_{a}+s_{b}-\frac{1}{2})C_{N}^{N/2}  \label{GN}
\end{equation}

As a consequence the residual entropy per triangle is
$\mathcal{S}=\ln 2$ at zero temperature.

The magnetization curve calculated with partition function
accounting only ground state manifold in the limit $h\gg T/N$ has
a form \cite{DK}
\begin{equation}
M=[s_{a}+s_{b}-\frac{1}{1+e^{h/T}}]N  \label{m3}
\end{equation}

As follows from Eq.(\ref{m3}) the magnetization per triangle
$m=\frac{M}{N}$ at $\frac{h}{T}\to 0$ tends to
$m=(s_{a}+s_{b}-\frac{1}{2})$. Therefore, the F-AF delta chain is
magnetically ordered at zero temperature in the critical point and
the magnetization undergoes a jump from $m=(s_{a}+s_{b})$ in the
ferromagnetic phase to $m=(s_{a}+s_{b}-\frac{1}{2})$ at the
critical point.

As we noted before the critical point separates different ground
state phases. One of them is ferromagnetic at $\alpha <\alpha
_{c}$. The question arises about a nature of the ground state and
the magnetic properties of the F-AF delta-chain at $\alpha >\alpha
_{c}$ and their dependence on spin values $s_{a}$ and $s_{b}$. For
example, the $s=\frac{1}{2}$ F-AF delta chain has the
ferrimagnetic ground state in the whole region $\alpha >\alpha
_{c} $ as it was shown in \cite{s=1/2,Rausch} on the base of
numerical calculations. But the structure of this state is not
quite clear especially for $\alpha \gg 1$.

In this paper we consider another example of the F-AF delta-chain consisting
of spins with $s_{a}=\frac{1}{2}$ and $s_{b}=1$. As it will be seen from the
following this model has more complex ground state phase diagram at $\alpha
>\alpha_c$ ($\alpha _{c}=\frac{1}{4}$) in contrast with the $s=\frac{1}{2}$
F-AF model. It is interesting to note that considered delta-chain
is closely related to another model with competing $F$ and $AF$
interactions, which is kagome-like spin-$\frac{1}{2}$ chain shown
in Fig.\ref{Fig_saw}b. The Hamiltonian of the latter has a form
\begin{equation}
\hat{H}=J_{1}\sum_{i=1}^{N}(\mathbf{\tau}_{i}+\mathbf{\xi
}_{i})\cdot (\mathbf{\sigma }_{i-1}+\mathbf{\sigma }_{i})
+J_{2}\sum_{i=1}^{N}(\mathbf{\tau}_{i}+\mathbf{\xi }_{i})\cdot
(\mathbf{\tau}_{i+1}+\mathbf{\xi }_{i+1}) \label{kagome}
\end{equation}%
where $\mathbf{\sigma }_{i}$, $\mathbf{\tau}_{i}$ and $\mathbf{\xi
}_{i}$ are $s=\frac{1}{2}$ operators of spins on axis, upper and
lower legs, respectively (see Fig.\ref{Fig_saw}b). The axis-leg
interaction $J_{1}$ is ferromagnetic and leg-leg interaction
$J_{2}$ is antiferromagnetic. (We further put $J_{1}=-1$ and
$J_{2}=\alpha $).

Model (\ref{kagome}) describes an interesting class of
quasi-one-dimensional compounds $Ba_{3}Cu_{3}In_{4}O_{12}$ and
$Ba_{3}Cu_{3}Sc_{4}O_{12}$ \cite{Bert,Maslova,Kumar}.

\begin{figure}[tbp]
\includegraphics[width=5in,angle=0]{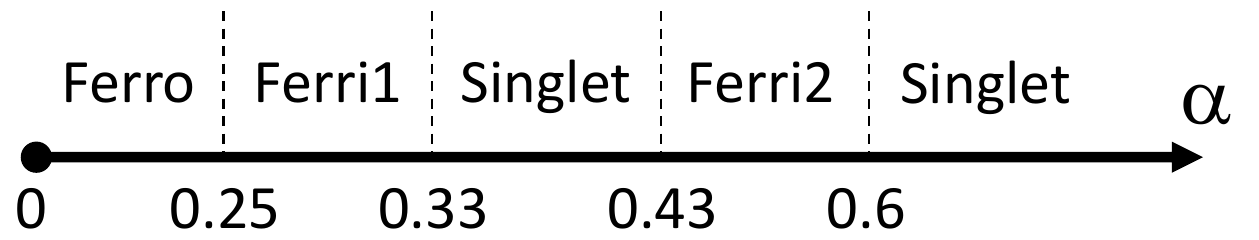}
\caption{The phase diagram of delta and kagome chain models.}
\label{Fig_phase}
\end{figure}

It is remarkable that the ground state phase diagram of the F-AF
delta-chain, with $s_{a}=\frac{1}{2}$, $s_{b}=1$ is the same as
that for the F-AF kagome chain (\ref{kagome}). This conclusion
follows from the comparison of basis states of both models. The
spins $\tau_{i}$ and $\xi _{i}$ on $i$-th sites of kagome legs are
incorporated into the compound spin ($\tau_{i}+\xi _{i}$) which is
either $1$, or $0$. Thus, the kagome chain (\ref{kagome}) is
equivalent to such delta-chain in which each basal site can be
occupied by spin $1$ or spin $0$. The site with spin $0$ is really
empty (defect) site because its spin does not interact with
neighbor basal spins $1$ and apical spins $\frac{1}{2}$, i.e. such
sites separate the delta-chain into an ensemble of finite chain
fragments decoupled from each other. It is easy to check that the
state with the lowest energy for given $S$ does not contain spin
$0$ sites, because each defect increases the energy by the value
$\Delta E\simeq 2|\varepsilon \left( S\right)|$, where
$\varepsilon \left( S\right) $ is the ground state energy per
triangle. Therefore, the ground states of the kagome chain
(\ref{kagome}) and the F-AF delta chain with $s_{a}=\frac{1}{2}$
and $s_{b}=1$ coincide in each spin sector for all values of
$\alpha $ and the magnetization at $T=0$ is the same for both
models. As it was shown in \cite{ferri} the ground state
degeneracy at the critical point $\alpha _{c}=\frac{1}{4}$ is the
same for both models and it is given by (\ref{GN}). The ground
state phase diagram of the kagome chain (\ref{kagome}) was
determined by T.\ Yamaguchi, Y.\ Ohta, and S.\ Nishimoto in
Ref.\cite{nishimoto}. According to results of \cite{nishimoto}
both models have five ground state phases as it is shown in
Fig.\ref{Fig_phase}. At $\alpha <\frac{1}{4}$ the ground state is
ferromagnetic and the magnetization undergoes a jump from
$m=\frac{3}{2}$ in the ferromagnetic phase to $m=1$ at the
critical point $\alpha_{c}$. The phases in regions
$0.25<\alpha<0.33$ and $0.42<\alpha <0.6$ are ferrimagnetic but
with different ground state spins, and the ground state is the
singlet in regions $0.33<\alpha<0.42$ and $\alpha>0.6$.

In the present paper we consider the magnetic properties in each
phase and study the zero-temperature magnetization processes in
them. We will show that the properties of the considered models at
$\alpha >\alpha _{c}$ are essentially different from those for the
$s=\frac{1}{2}$ delta-chain. In particular, this concerns the
region of large $\alpha $ where these models have different ground
states and different behavior of the magnetization. In order to
study the ground state and the spectrum of the delta-chain with
$s_{a}=\frac{1}{2}$ and $s_{b}=1$ we employ analytical approaches
and exact diagonalization (ED) and DMRG numerical calculations.

The paper is organized as follows. In Section II we consider the
magnetic properties of the F-AF model in the ferrimagnetic phase
Ferri1. We give an analytical estimate of the magnetization $m(h)$
and compare it with numerical results. In Section III we study the
magnetization in an intermediate singlet phase and show that jumps
in the magnetization curve occur in this phase. In Section IV we
consider the ferrimagneic phase Ferri 2 in which the magnetization
curve has a plateau. In Section V we study the magnetization in
the singlet phase at $\alpha >0.6$. In the region $0.6<\alpha <2$
the magnetization jump occurs and at $\alpha>2$ the magnetization
curve has a plateau. In this section we also compare the behavior
of the considered models with that for the F-AF $s=\frac{1}{2}$
delta-chain at $\alpha\gg 1$. In Section VI we give a summary.

\section{Vicinity of the critical point $0.25<\alpha <0.33$}

For $\alpha >\frac{1}{4}$ the macroscopic ground state degeneracy
(\ref{GN}) splits. At small value of $\gamma $ ($\gamma =\alpha
-\frac{1}{4}$) the total spin of the ground state $S_{tot}$ is
close to $N$. The numerical calculations of finite chains give
$S_{tot}=N$ for $N<12$, $S_{tot}=N+1$ for $12\leq N<30$ and
$S_{tot}=N+2$ for $N=30$ at $\gamma =0.01$. It is unclear from
these numerical data whether the excess over $S_{tot}=N$ is
thermodynamic or it vanishes in the thermodynamic limit. To answer
this question and determine the value of the total spin of the
ground state in the thermodynamic limit, we study the behavior of
the magnetization curve, $m\left( h\right) $.

At first we consider the behavior of $m\left( h\right) $ near the
saturation field $h=h_{s}$. Above the saturation field the system
is, ferromagnetically ordered. Below $h_{s}$ the model is
described in terms of down spins (magnons) on the ferromagnetic
background. The saturation field is equal to a minimal energy of
the one-magnon state which is $E_{1}=-4\gamma $, i.e.
$h_{s}=4\gamma $.

The system of magnons near $h_{s}$ can be considered as the
Bose-gas with the $\delta $-function interaction, which at low
density is equivalent to spinless fermions
\cite{Sakai,Okunishi,Affleck}. Spin operators of basal spin-$1$
$\mathbf{S}_{n}$ and apical spin-$\frac{1}{2}$ $\mathbf{\sigma
}_{n} $ in the fermion representation are:
\begin{eqnarray}
S_{n}^{+} &=&\sqrt{2}b_{n},\quad S_{n}^{z}=1-b_{n}^{+}b_{n} \\
\sigma _{n}^{+} &=&a_{n},\quad \sigma _{n}^{z}=\frac{1}{2}-a_{n}^{+}a_{n}
\end{eqnarray}

Then the Hamiltonian takes the form:%
\begin{eqnarray}
\hat{H}_{f} &=&E_{F}-\frac{1}{\sqrt{2}}%
\sum_{n=1}^{N}[b_{n}^{+}(a_{n}+a_{n-1})+(a_{n}^{+}+a_{n-1}^{+})b_{n}]+
\nonumber \\
&&+\alpha \sum_{n=1}^{N}(b_{n}^{+}b_{n+1}+b_{n+1}^{+}b_{n})+(1-2\alpha
)\sum_{n=1}^{N}b_{n}^{+}b_{n}+2\sum_{n=1}^{N}a_{n}^{+}a_{n}
\end{eqnarray}%
where $E_{F}=(\alpha -1)N$ is the energy of the ferromagnetic
state. Near $h_{s}$ the number of fermions is small and we ignored
in $\hat{H}_{f}$ four-fermion terms.

Diagonalizing the Hamiltonian $\hat{H}_{f}$, we arrive at%
\begin{equation}
\hat{H}_{f}=E_{F}+\sum_{k}E_{-}(k)A_{k}^{+}A_{k}+%
\sum_{k}E_{+}(k)B_{k}^{+}B_{k}
\end{equation}%
where $A_{k}$ and $B_{k}$ are new Fermi-operators and the spectrum
$E_{\pm }(k)$ is
\begin{equation} E_{\pm }(k)=\frac{3}{2}-\alpha
(1-\cos k)\pm \lbrack (\frac{1}{2}+\alpha (1-\cos k))^{2}+(1+\cos
k)]^{1/2}
\end{equation}

$E_{-}(k)$ is the lowest branch and its minimum is $E_{-}(k=\pi
)=-4\gamma $. The $z$ -projection of total spin $S^{z}$ is given by%
\begin{equation}
S^{z}=\frac{3N}{2}-\sum_{k}(A_{k}^{+}A_{k}+B_{k}^{+}B_{k})
\end{equation}

At magnetic field close to the saturated value, the number of
fermions is small and we fill the lowest branch $E_{-}(k)$ in the
momentum space region $[\pi -k_{F},\pi +k_{F}]$. Then the energy
in the magnetic field $h$ becomes
\begin{equation}
E=E_{F}+\frac{N}{\pi }\int_{\pi -k_{F}}^{\pi }E_{-}(k)dk-hS^{z}
\end{equation}%
where
\begin{equation}
S^{z}=\frac{3N}{2}-N\frac{k_{F}}{\pi }
\end{equation}

The dependence $m(h)$ is obtained from the condition
$\frac{dE}{dk_{F}}=0$, which leads to the equation for $k_{F}(h)$:
\begin{equation}
E_{-}(\pi -k_{F})+h=0
\end{equation}

When $h$\ is close to $h_{s}$, the value of $k_{F}\ll 1$ and the leading
term for $m(h)$ is%
\begin{equation}
m=\frac{3}{2}-\frac{2\sqrt{2}\sqrt{1+2\gamma }}{\pi \sqrt{3+4\gamma }}\sqrt{%
1-\frac{h}{h_{s}}}  \label{mhg}
\end{equation}

As it is known \cite{Hodgeson,Akutsu} the magnetization curve near
$h_{s}$ can be constructed with the use of the solution of
two-magnon problem. Our consideration of the state with
$S^{z}=\frac{3N}{2}-2$ shows that the leading term of $m(h)$ for
$(h_{s}-h)\ll h_{s}$ coincides with Eq.(\ref{mhg}).

If $\gamma \ll 1$ then $m(h)$ is%
\begin{equation}
m=\frac{3}{2}-b\sqrt{1-\frac{h}{h_{s}}}  \label{mh}
\end{equation}%
where $b=\frac{2}{\pi }\sqrt{\frac{2}{3}}\approx 0.52$.

\begin{figure}[tbp]
\includegraphics[width=5in,angle=0]{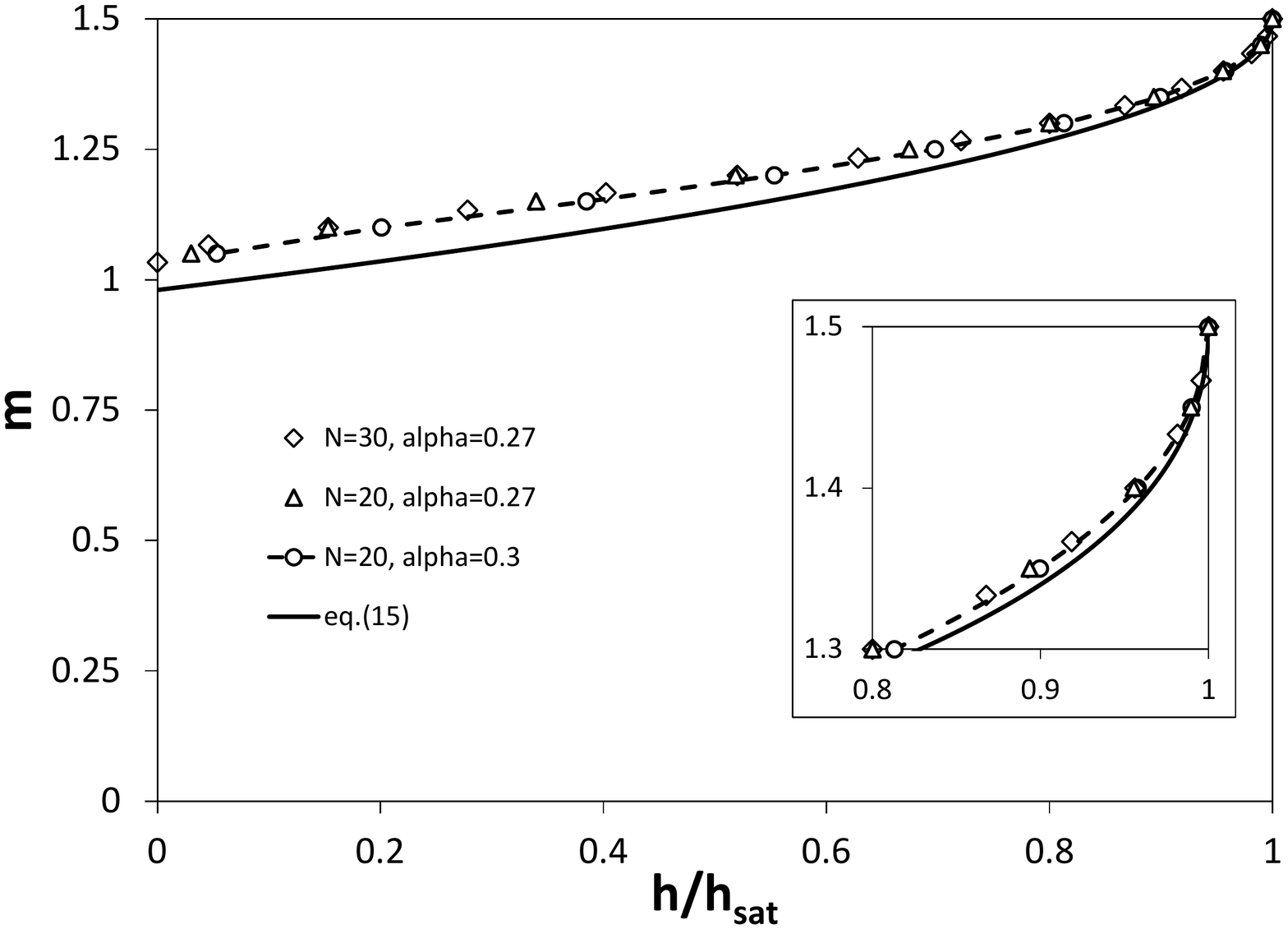}
\caption{Magnetization curve near the transition point. The inset
shows the behavior of the magnetization curve near the saturation
field.} \label{Fig_M_h_a027}
\end{figure}

Though Eq.(\ref{mh}) is valid for $h$ close to $h_{s}$, it turns
out that this estimate gives reasonable accuracy up to the case
$h=0$. In Fig.\ref{Fig_M_h_a027} numerical calculations (DMRG) of
the magnetization curve for $N=20,30$ and $\alpha =0.27,0.3$ are
compared with Eq.(\ref{mh}). As it is seen on the inset of
Fig.\ref{Fig_M_h_a027}, Eq.(\ref{mh}) perfectly describes the
numerical results near the saturation field. For low and
intermediate values of the magnetic field, the analytical estimate
(\ref{mh}) slightly diverges from the numerical results and,
finally, Eq.(\ref{mh}) predicts the total spin of the ground state
$S_{tot}\simeq 0.98N$ at $h=0$. At the same time, the numerical
calculations shown in Fig.\ref{Fig_M_h_a027} indicate that all
magnetization curves for different $N$ are very close to each
other and all of them tends to the same value $S_{tot}\simeq
1.04N$ at $h\to 0$. Thus, the small excess of the total spin of
the ground state over $S_{tot}=N$ detected on finite chains does
not vanish in the thermodynamic limit and leads to an increase of
$S_{tot}$ by $4\%$.

The case when all spins of the delta-chain (both apical and basal)
are $s_{a}=s_{b}=\frac{1}{2}$ can be also studied in terms of the
fermion representation of magnons. The leading term for the
magnetization $m(h)$ near the saturation field $h_{s}=2\gamma$
($\gamma =\alpha -\frac{1}{2}$)
has a form%
\begin{equation}
m=1-\frac{\sqrt{2}\sqrt{1+2\gamma }}{\pi \sqrt{1+\gamma }}\sqrt{1-\frac{h}{%
h_{s}}}  \label{mh12}
\end{equation}

Eq.(\ref{mh12}) gives the estimate for the total spin of the
ground state $S_{tot}\simeq 0.55N$ ($\gamma \ll 1$) at $h=0$,
which is in very good agreement with the ground state spin
$S_{tot}\simeq 0.54N$ obtained by numerical calculations
\cite{Rausch}. Therefore, the value of $S_{tot}\simeq 0.54N$ in
the $s_{a}=s_{b}=\frac{1}{2}$ delta chain model slightly exceeds
the value of magnetization $S_{tot}=0.5N$ at the critical point.
This fact is in accord with the studied models, where the ground
state total spin $S_{tot}\simeq 1.04N$ is slightly above the value
of the magnetization at the critical point $M=N$.

\section{Intermediate singlet phase $0.33<\alpha <0.43$}

\begin{figure}[tbp]
\includegraphics[width=5in,angle=0]{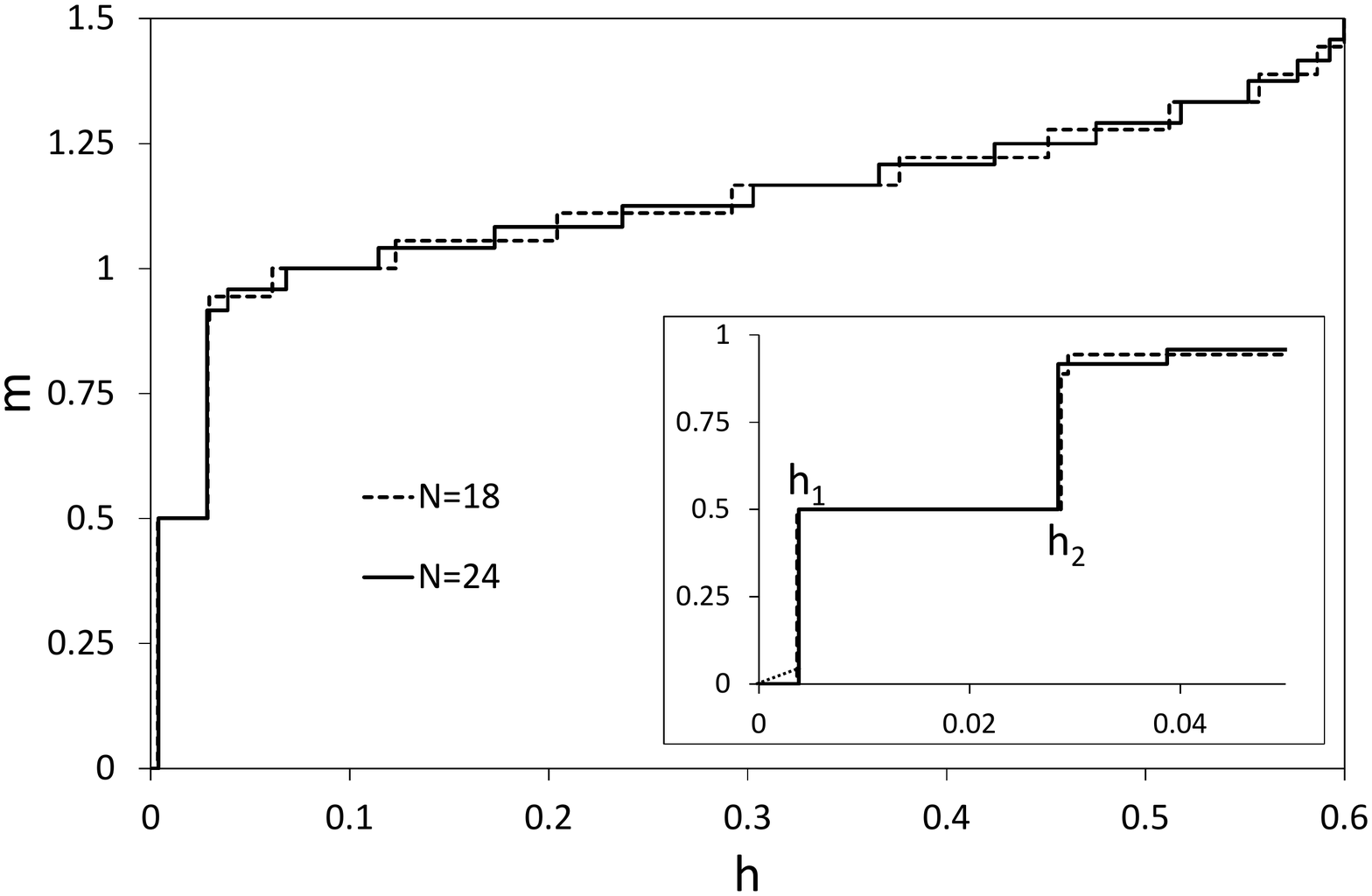}
\caption{Magnetization curve for $\alpha =0.4$ and $N=18,24$. The
inset shows the low magnetic field part of the magnetization
curve.} \label{Fig_M_h_a04}
\end{figure}

Intermediate singlet phase is realized in the region $0.33<\alpha
<0.43$. The singlet ground state has a period $6$
\cite{nishimoto}, so that the periodic chains of length $N=6k$
have singlet ground states, while systems of other lengths have
ground states with non-zero total spin $S_{tot}$, which is however
vanishing in the thermodynamic limit $N\to\infty$. The
magnetization curve for $\alpha =0.4$ is shown in
Fig.\ref{Fig_M_h_a04} for $N=18,24$. As it is seen in
Fig.\ref{Fig_M_h_a04} the magnetization curve has two
magnetization jumps, the first one occurs at $h_{1}\simeq 0.0036$
from the singlet state to $S_{tot}=N/2$, then after magnetization
plateau at $m=\frac{1}{2}$, the second jump takes place at
$h_{2}\simeq 0.0285$ from $m=\frac{1}{2}$ to the value slightly
lower than $m=1$.

Both fields $h_{1}$ and $h_{2}$ tend to zero when $\alpha
\rightarrow 0.33$, providing the phase transition to the
ferrimagnetic ground state with $S_{tot}\simeq 1.04N$. On the
other side of the region, $\alpha \rightarrow 0.43$, the
transition to another ferrimagnetic phase with $S_{tot}=N/2$ takes
place, as will be shown in the next section. Therefore, the first
field $h_{1}$ vanishes at the phase boundary $\alpha =0.43$, while
the second magnetization jump continuously transform to the steep
square-root behavior of $m(h)$ above $m=\frac{1}{2}$ in the
ferrimagnetic region $0.43<\alpha <0.6$.

The plateau at $m=0$ shown in Fig.\ref{Fig_M_h_a04} for $N=18,24$
at first glance contradicts to the numerical results of zero
singlet-triplet energy gap presented in Fig.5 of
Ref.\cite{nishimoto}. However, this is just finite-size effect,
that is the value of $h_{1}\simeq 0.0036$ is so low, that to
observe the first step on the magnetization curve (at $h$ equal to
the singlet-triplet gap) one need to calculate the chains much
longer than $N=24$. So that in the thermodynamic limit there is an
approximately linear growth of magnetization in the region
$0<h<h_{1}$, which we indicated by dotted line in
Fig.\ref{Fig_M_h_a04}.

\section{Ferrimagnetic phase $0.43<\alpha <0.6$}

\begin{figure}[tbp]
\includegraphics[width=5in,angle=0]{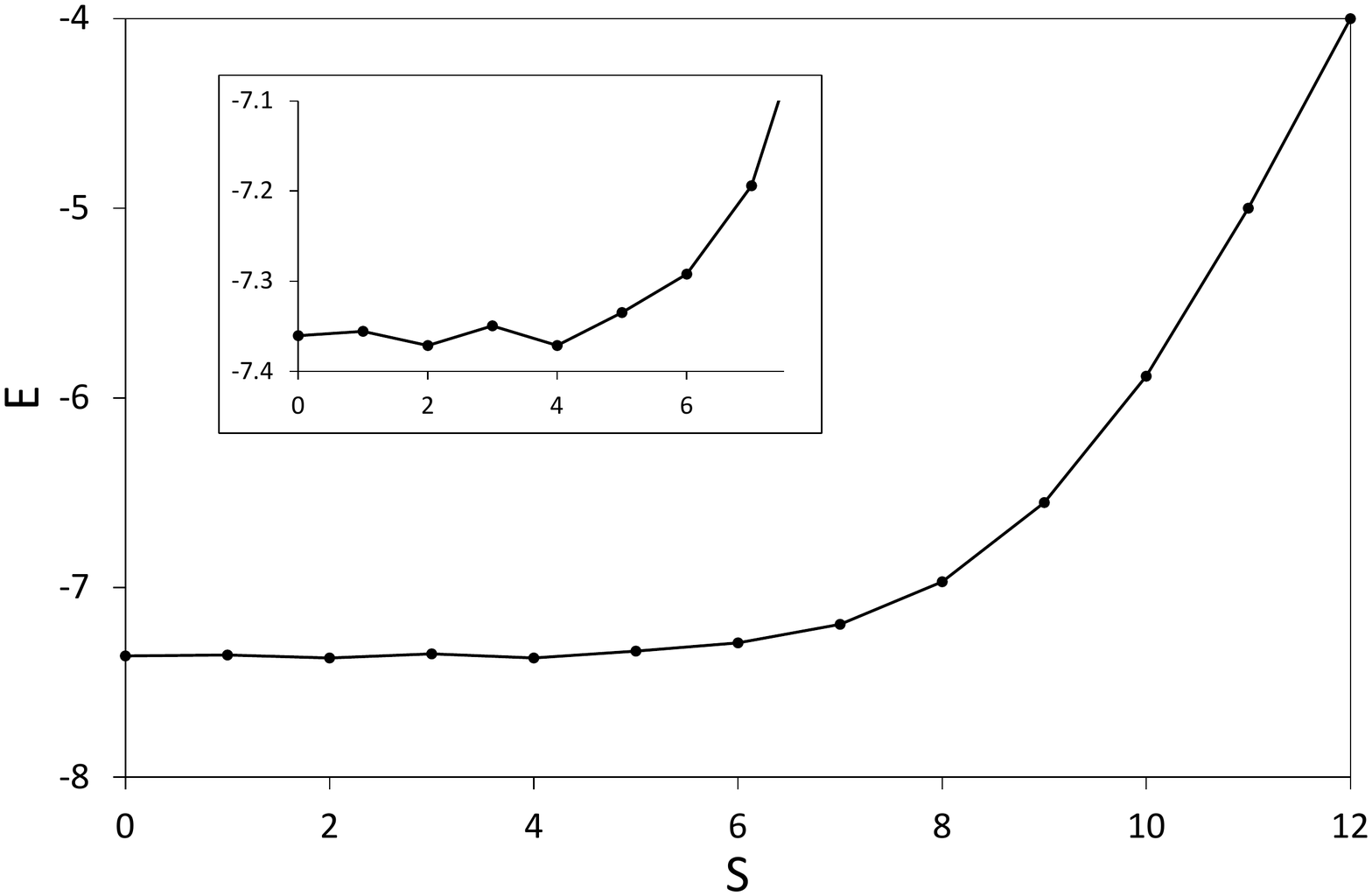}
\caption{Spectrum $E(S)$ for $\alpha =0.5$ and $N=8$. The inset
shows the low energy part of $E(S)$.} \label{Fig_ES_a05_N8}
\end{figure}

When $\alpha >0.43$, the total spin of the ground state changes
from $S_{tot}=0$ to $S_{tot}=N/2$ \cite{nishimoto}. The dependence
$E(S_{tot})$ in this region is very flat for $S_{tot}\leq N/2$, so
that the energy difference is negligible in the thermodynamic
limit, and ground state is quasi-degenerate with $0\leq
S_{tot}\leq N/2$. The example of the dependence $E(S_{tot})$ is
presented in Fig.\ref{Fig_ES_a05_N8} for $N=8$.

\begin{figure}[tbp]
\includegraphics[width=5in,angle=0]{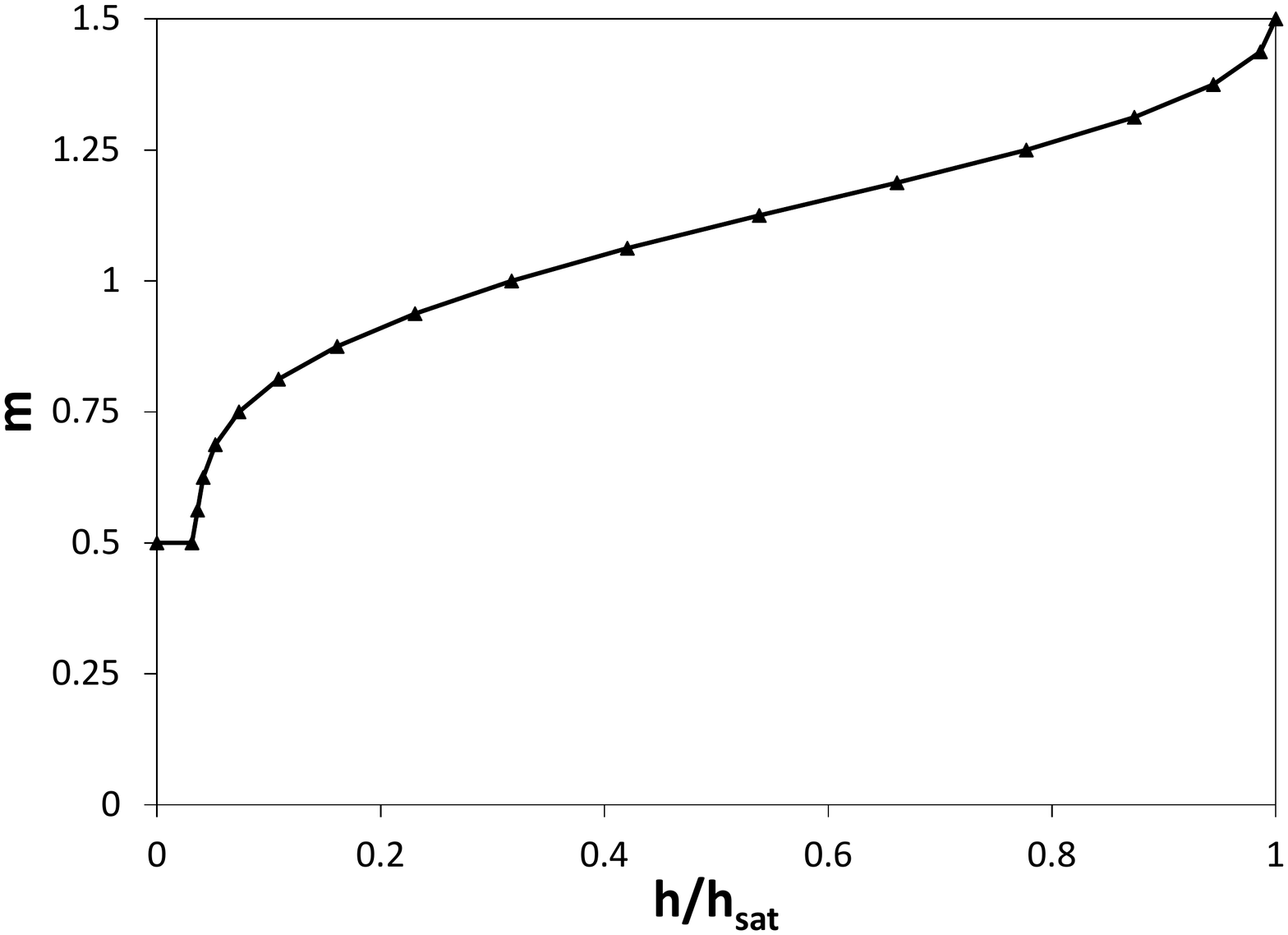}
\caption{Magnetization curve for $\alpha =0.5$ and $N=16$.}
\label{Fig_M_h_a05}
\end{figure}

The ferrimagnetic phase in this region is accompanied by the
energy gap for the states with $S_{tot}=N/2+1$, which manifests
itself in the plateau at $m=\frac{1}{2}$ on the magnetization
curve. The numerical data confirms these facts. The magnetization
curve for $N=16$ at $\alpha =0.5$ is shown in
Fig.\ref{Fig_M_h_a05}. One can see the plateau at $m=\frac{1}{2}$
up to the magnetic field $h_{up}\simeq 0.031$, which is in accord
with the numerical data for the energy gap shown in Fig.2 of
Ref.\cite{nishimoto}. Above $h_{up} $ the magnetization curve
shows square-root-like behavior $m-\frac{1}{2}\sim
\sqrt{h-h_{up}}$. Similar square-root behavior occurs near the
saturation field, where $\frac{3}{2}-m\sim \sqrt{h_{s}-h}$.

The case $\alpha =0.5$ is a generic case for the region
$0.43<\alpha <0.6$, so that the behavior of the system in the
whole region is similar to the case $\alpha =0.5$.

\section{Singlet phase $\alpha>0.6$}

We start to study the singlet phase $\alpha>0.6$ from the limit
$\alpha \gg 1 $. When $\alpha =\infty$ the model consists of two
non-interacting subsystems: the basal antiferromagnetic chain and
isolated apical spins. For large but finite $\alpha $ the
interaction between two subsystems can be treated in the frame of
the perturbation theory (PT) in small parameter $\frac{1}{\alpha
}$. In this limit it is convenient to renormalize the interactions
as $J_{2}=1$ and $J_{1}=-\frac{1}{\alpha }$ and the Hamiltonian
(\ref{H}) takes the form%
\begin{eqnarray}
\hat{H} &=&\hat{H}_{0}+\hat{V} \\
\hat{H}_{0} &=&\sum \mathbf{S}_{i}\cdot \mathbf{S}_{i+1} \\
\hat{V} &=&-\frac{1}{\alpha }\sum (\mathbf{S}_{i-1}+\mathbf{S}_{i})\cdot
\mathbf{\sigma }_{i}
\end{eqnarray}%
where $\hat{H}_{0}$ is the Hamiltonian of the antiferromagnetic
basal chain and $\hat{V}$ is the basal-apical interaction.

The ground state of $\hat{H}_{0}$ is a singlet and by its symmetry
the first order in $\hat{V}$ is zero. However, in the triplet
state of basal subsystem the first order of PT is nonzero and
$\left\langle \hat{V}\right\rangle $ is
\begin{equation}
\left\langle \hat{V}\right\rangle =-\frac{1}{\alpha }\sum \left\langle
S_{i}^{z}+S_{i+1}^{z}\right\rangle \sigma _{i}^{z}  \label{V}
\end{equation}

The average $\left\langle S_{i}^{z}+S_{i+1}^{z}\right\rangle $ in
the triplet state is $\left\langle
S_{i}^{z}+S_{i+1}^{z}\right\rangle =\frac{2}{N}$ (PBC is assumed)
and $\left\langle \hat{V}\right\rangle $ has the minimum when all
$\sigma _{i}^{z}$ $=\frac{1}{2}$, i.e. the apical subsystem is
fully polarized, so that the total spin of the system is
$S_{tot}=\frac{N}{2}+1$ ($S=S_{a}+S_{b}$). The energy of the
lowest triplet of $\hat{H}_{0}$ (the singlet-triplet gap) is
$\Delta _{1}=0.4105$ \cite{White}. Therefore, in the first order
of PT the energy of the state with $S_{tot}=\frac{N}{2}+1$ (the
basal triplet and the fully polarized apical subsystem) is
\begin{equation}
E_{1}=E_{0}+(\Delta _{1}-\frac{1}{\alpha })  \label{Etriplet}
\end{equation}%
where $E_{0}=-1.41N$ \cite{White}. Therefore, for $\alpha \gg 1$
the state with $S_{tot}=\frac{N}{2}+1$ can not be the ground
state, because the small apical-basal interaction can not overcome
the finite singlet-triplet gap in the basal subsystem.

As we noted before, the first order PT in $\frac{1}{\alpha }$ for
the singlet ground state of $\hat{H}_{0}$ is zero and, therefore,
independent of the apical subsystem configuration. In order to
determine the ground state of the apical subsystem we need to
study the next order of PT. The second order PT for the singlet
ground state of the basal subsystem leads to an effective
spin-$\frac{1}{2}$ Hamiltonian for the apical subsystem, as it was
shown in Ref.\cite{Chandra}. Though the basal-apical interaction
in the model studied in Ref.\cite{Chandra} is AF, the sign of this
interaction is unimportant in the second order PT in
$\frac{1}{\alpha }$ and the results of \cite{Chandra} are
applicable to the present model. The effective Hamiltonian
describing the apical subsystem is rather unusual. It represents
two weakly interacting and frustrated spin-$\frac{1}{2}$
Heisenberg spin-chains on odd and even apical sites. The ground
state of this effective Hamiltonian is the singlet and the lowest
energy $E(S_{a})$ of the state with the total apical spin $S_{a}$
is a smoothly increasing function up to $S_{a}=\frac{N}{2}$ with a
small prefactor $\alpha ^{-2}$.

Based on the above we can describe the behavior of the system in
the external magnetic field. At zero magnetic field the ground
state of the model is the singlet consisting of both basal and
apical singlets. For very low magnetic fields $h\sim \alpha ^{-2}$
the apical subsystem becomes partly polarized, while the basal
subsystem is in its singlet state. The magnetization per triangle
$m=S_{tot}/N$ smoothly increases with $h$ and the dependence
$m(h)$ is governed by the function $E(S_{a})$. At some "apical
saturation field" $h_{low}$, the apical subsystem becomes fully
polarized and $m=\frac{1}{2}$. Using the parameters of the
effective Hamiltonian given in \cite{Chandra} we estimate this
field as $h_{low}\approx 0.26/\alpha ^{2}$. The magnetization
curve $m(h)$ is steadily increases in the region $0\leq h\leq
h_{low}$ and $m(h)$ near $h_{low}$ behaves as $m(h)\sim
\frac{1}{2}-\sqrt{1-\frac{h}{h_{low}}}$.

Further increase of the magnetic field $h>h_{low}$ does not lead
to increase of the magnetization, providing a magnetization
plateau at $m=\frac{1}{2}$, untill the magnetic field overcomes
the singlet-triplet gap of the basal subsystem. The state with
$S=\frac{N}{2}+1$ is represented in the first order PT as the
lowest basal triplet in the field $h+\frac{1}{\alpha }$ formed by
the polarized apical spins and external magnetic field. The energy
of this state is given by Eq.(\ref{Etriplet}) and it determines
the field $h_{up}=(\Delta _{1}-\frac{1}{\alpha })$ at which the
magnetization starts to increase from the plateau at
$m=\frac{1}{2}$. The difference $h_{up}-h_{low}$ is a width of the
plateau of the magnetization at $m=\frac{1}{2}$. The states with
$S>\frac{N}{2}+1$ can be considered in the first order PT as the
spin-$1$ Heisenberg basal chain in the magnetic field
$h+\frac{1}{\alpha }$. The behavior of $E(S)$ near $S\geq
\frac{N}{2}$ is the same as that for the
spin-$1$ Heisenberg chain \cite{Affleck}%
\begin{equation}
E(S_{b})=\Delta _{1}S_{b}+d\frac{S_{b}^{3}}{N^{2}}
\end{equation}%
where $d\simeq 25$. This leads to the following behavior of the
magnetization curve $m(h)$ near $h_{up}$:
\begin{equation}
m=\frac{1}{2}+0.115\sqrt{h-h_{up}}
\end{equation}

Further strengthening of the magnetic field from $h_{up}$ to
$h_{sat}=4\gamma $ results in the continuous increase of the
magnetization and near $h_{sat}$ the magnetization behaves as
\begin{equation}
m=\frac{3}{2}-\frac{4}{\pi }\sqrt{1-\frac{h}{h_{sat}}}
\end{equation}

\begin{figure}[tbp]
\includegraphics[width=5in,angle=0]{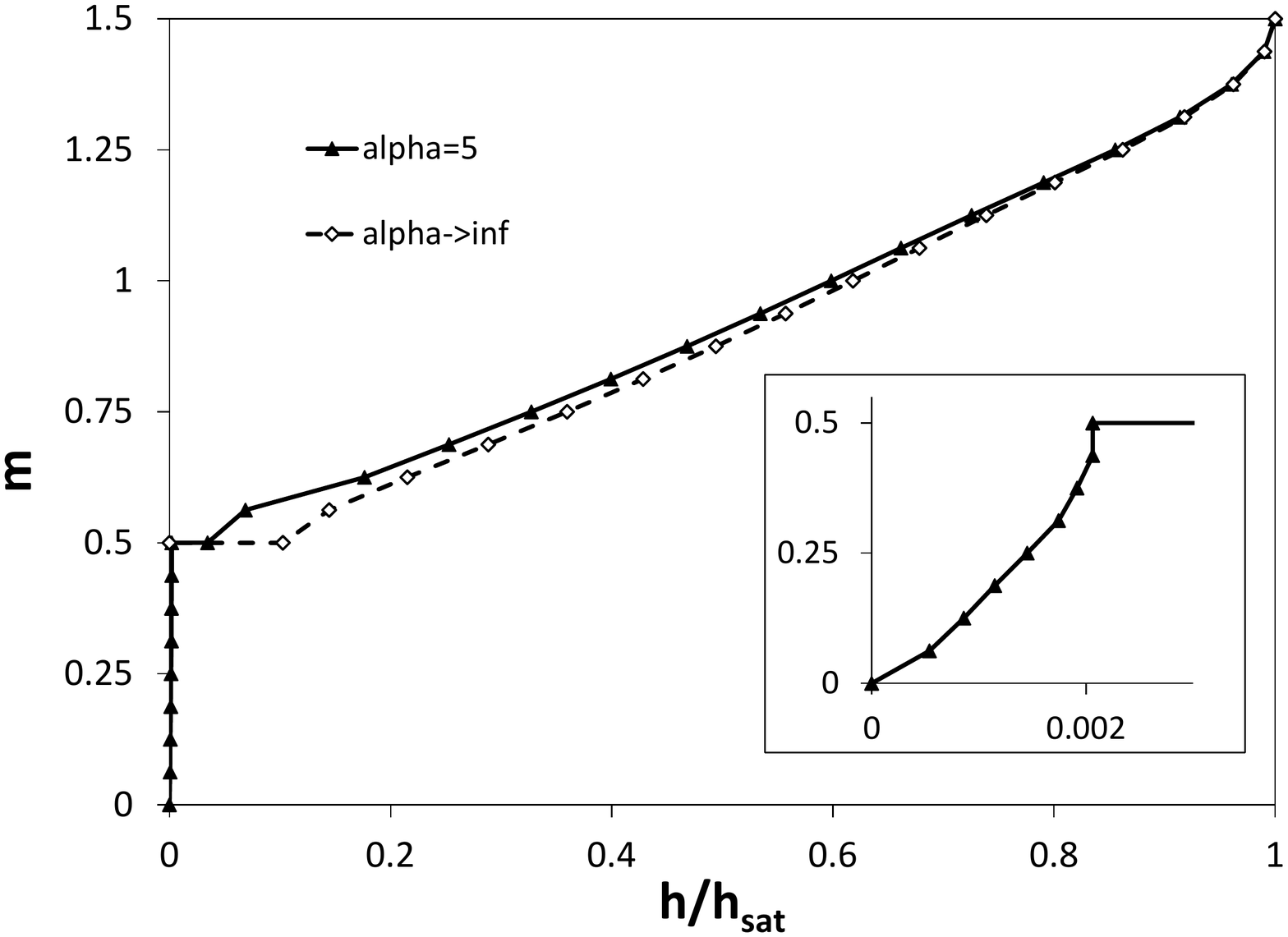}
\caption{Magnetization curve for $\alpha =5$ and $N=16$. The inset
shows the low magnetic field part of the magnetization curve.}
\label{Fig_M_h_a5}
\end{figure}

The numerical calculations confirm the above reasoning. The
magnetization curve for $\alpha =5$ is shown in
Fig.\ref{Fig_M_h_a5} in comparison with the case $\alpha \to
\infty $. As it is seen in Fig.\ref{Fig_M_h_a5} both magnetization
curves have plateau at $m=\frac{1}{2}$. In the inset of
Fig.\ref{Fig_M_h_a5} the behavior of the magnetization curve is
shown for low magnetic field, where the magnetization occurs on
the apical subsystem only.

\begin{figure}[tbp]
\includegraphics[width=5in,angle=0]{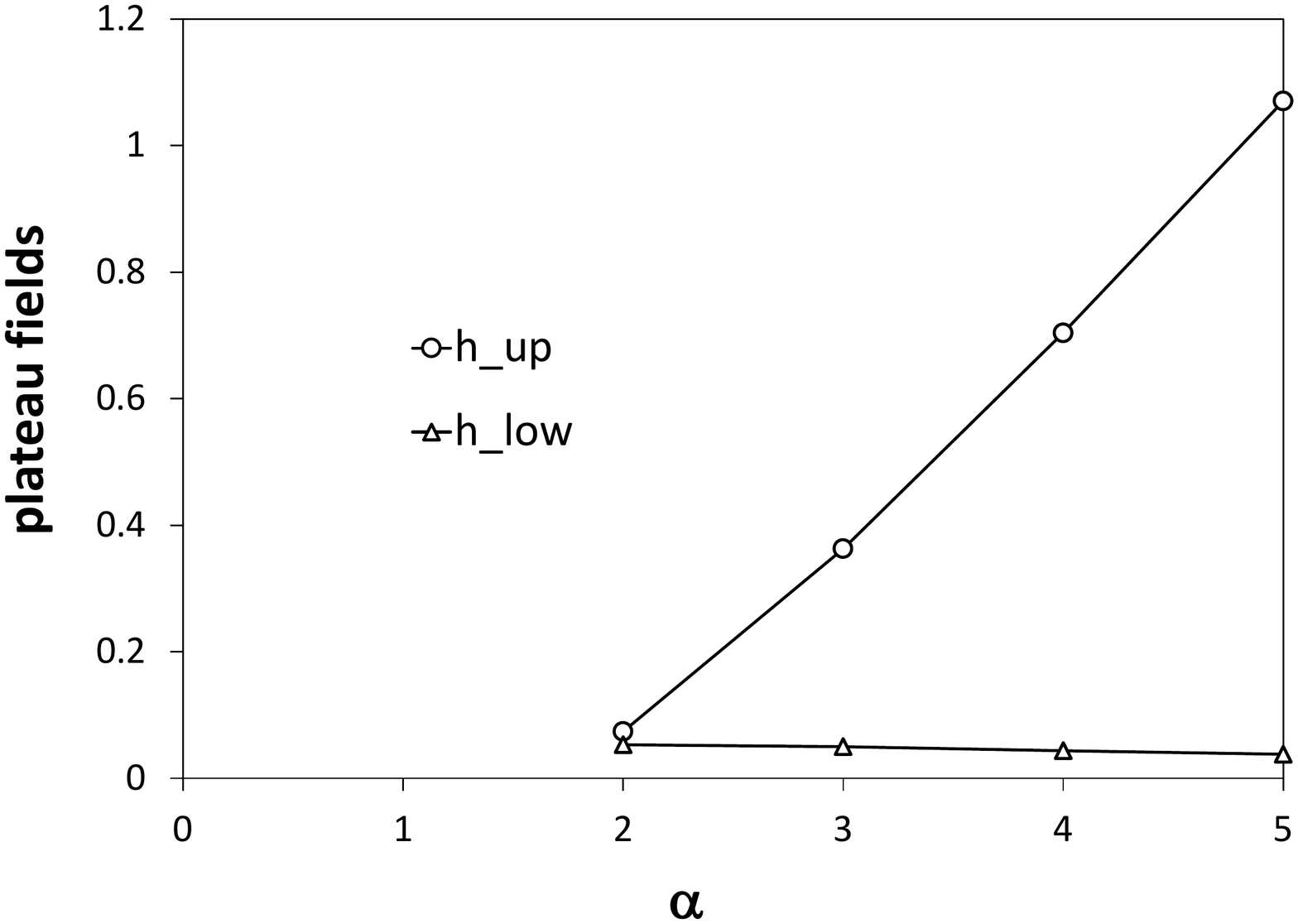}
\caption{Dependence of up and low magnetic fields, determining the
magnetization plateau, on $\alpha $.} \label{Fig_huhd}
\end{figure}

\begin{figure}[tbp]
\includegraphics[width=5in,angle=0]{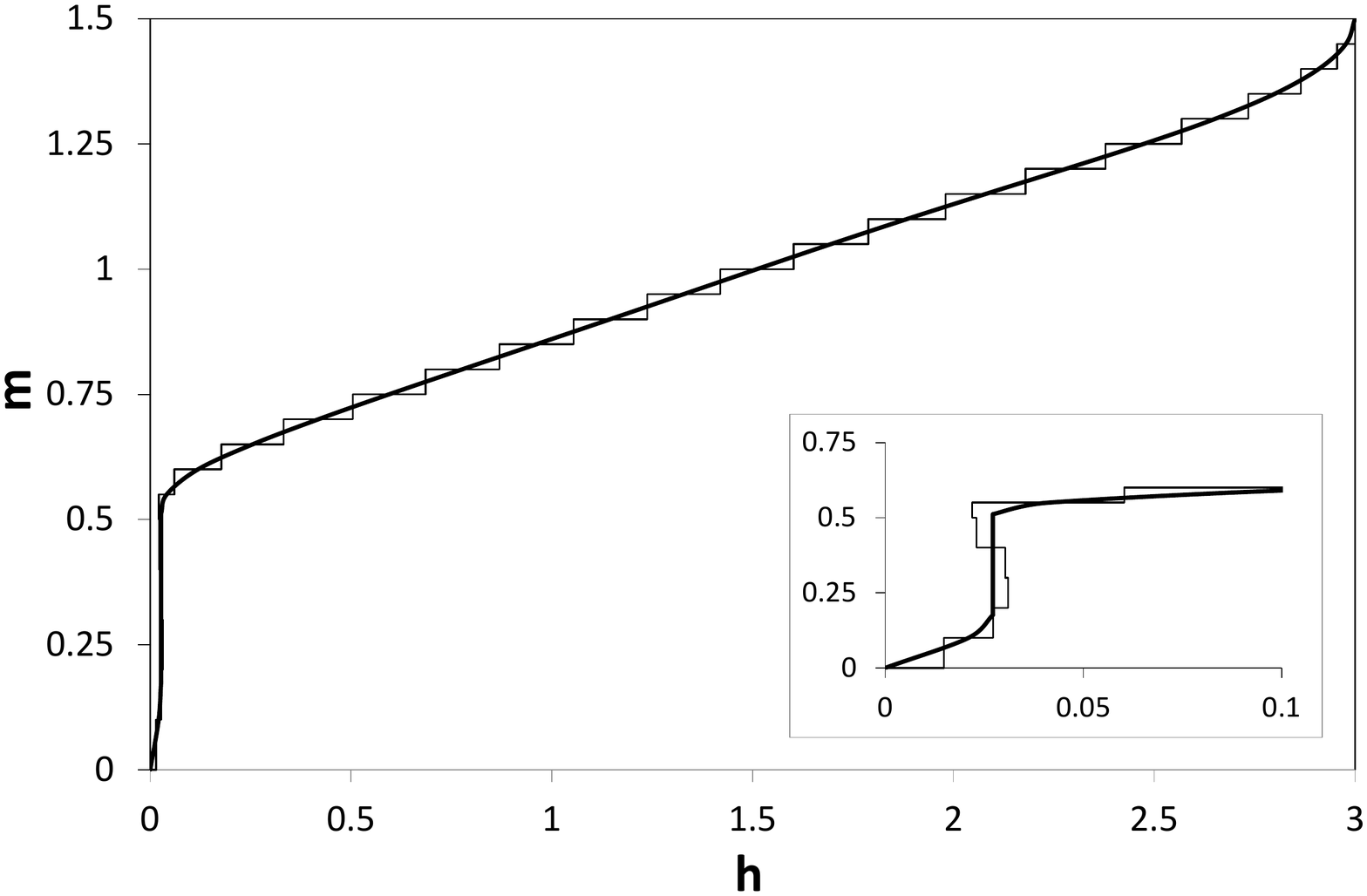}
\caption{Magnetization curve for $\alpha =1$ and $N=20$. The inset
shows the low magnetic field part of the magnetization curve. The
S-shaped form of the magnetization curve constructed as
$h(S)=E(S+1)-E(S)$ (thin lines) indicates the concavity of the
dependence $E(S)$ and the corresponding jump in magnetization
(thick line).} \label{Fig_M_h_a1}
\end{figure}

The dependencies of $h_{up}$ and $h_{low}$ on $\alpha $ are shown
in Fig.\ref{Fig_huhd}. As follows from Fig.\ref{Fig_huhd} the
plateau vanishes at some value of $\alpha $ slightly lower than
$\alpha =2$. It means that for $\alpha <2$ the above speculative
separation of the system on almost independent apical and basal
subsystems is no longer relevant. The behavior of the
magnetization curve in Fig.\ref{Fig_M_h_a1} for $\alpha =1$
indicates that after the vanishing of the plateau the metamagnetic
jump at $h=h_{jump}$ is forming from some finite value
$m_{1}<\frac{1}{2}$ to the value $m_{2}$ close to $m=\frac{1}{2}$.
When $\alpha$ decreases, $h_{jump}$ decreases also and
$h_{jump}\to 0$ when $\alpha \to 0.6$, so that the jump occurs
from $m_{1}\to 0$ to $m_{2}\to \frac{1}{2}$. Finally, at $\alpha
=0.6$ the transition from the singlet phase to the ferrimagnetic
phase with $m=\frac{1}{2}$ occurs.

Now let us compare the behavior of the considered models with that
of spin-$\frac{1}{2}$ F-AF delta-chain in the $\alpha \gg 1$
limit. The main difference stems from different spectra of the
basal AF chains: the gapped one for spin-$1$ and gapless for
spin-$\frac{1}{2}$. The energy of the lowest triplet of
$\hat{H}_{0}$ for the case $s_{b}=\frac{1}{2}$ is $\Delta E=\pi
^{2}/N$. In the first order of PT (\ref{V}) the energy difference
between the states with $S_{tot}=\frac{N}{2}+1$ (the basal triplet
and fully polarized apical subsystem) and $S_{tot}=0$ (the basal
singlet and non-magnetic apical subsystem) is
\begin{equation}
E(S=\frac{N}{2}+1)-E(S=0)=\frac{\pi ^{2}}{N}-\frac{1}{\alpha }
\end{equation}

So, the state with $S_{tot}=\frac{N}{2}+1$ has lower energy for
$N>\pi ^{2}\alpha $. The numerical value of $\pi ^{2}\alpha $ is
large, so one should be careful when interpreting the results of
numerical calculations of finite chains. For example, for the
delta-chain with $\alpha =10$ the magnetic state with
$S_{tot}=\frac{N}{2}+1$ is realized for $N>100$, while for $N<100$
one observe the singlet ground state.

If we study the delta-chain of macroscopical size, we should
examine the starting states of the basal subsystem with different
macroscopical value of the total spin $S_{b}\sim N$ and their
energies are \cite{Grif}
\begin{equation}
E_{b}(S)=E_{0b}+\frac{\pi ^{2}S_{b}^{2}}{2N}
\end{equation}%
where $E_{0b}=(\frac{1}{4}-\ln 2)N$.

Taking into account the first order of PT (\ref{V}) the energy of
the total system becomes
\begin{equation}
E(S)=E_{0b}+\frac{\pi ^{2}S_{b}^{2}}{2N}-\frac{S_{b}}{\alpha}
\label{ES12}
\end{equation}

Minimization of (\ref{ES12}) over $S_{b}$ gives $S_{b}=N/\pi
^{2}\alpha $. Thus, the ground state for large $\alpha $ is
ferrimagnetic with total spin $S_{tot}=N\left(
\frac{1}{2}+\frac{1}{\pi^{2}\alpha }\right) $.

The magnetization as a function of the external field $h$, rises
continuously from the value
\begin{equation}
m=\frac{1}{2}+\frac{1}{\pi^{2}\alpha } \label{ms}
\end{equation}%
at $h=0$ to $m=1$ at the saturation field $h_{s}=2\alpha -1$. We
stress that the obtained magnetization (\ref{ms}) exceeds the
value $m=\frac{1}{2}$ by a small, but finite value, describing the
weak induced polarization of the basal subsystem. Near the
saturation field the magnetization shows a common square root
behavior: $1-m\sim \sqrt{h_{s}-h}$.

Hence, the ground state of model (\ref{H}) with basal spin-$1$ and
spin-$\frac{1}{2}$ in the $\alpha \gg 1$ limit are different -
singlet ground state for basal spin-$1$ and ferrimagnetic ground
state with magnetization slightly higher than $m=\frac{1}{2}$ for
basal spin-$\frac{1}{2}$. We expect that such difference remains
for all other values of basal spins: singlet ground state for
integer values of basal spins and ferrimagnetic ground state for
half-integer values of basal spins.

\section{Summary}

We have studied the magnetic properties of the frustrated model
consisting of the triangles with competing ferromagnetic
interactions between basal spins-$1$ and apical spins-$\frac 12$
and antiferromagnetic interactions between basal spins. The
magnetic properties of this model at zero temperature exactly
coincides with that of the kagome-like spin-$\frac 12$ chain.

The ground state of these models depend on the ratio between
antiferromagnetic and ferromagnetic interactions, $\alpha$, and
contains 5 phases: the ferromagnetic phase $\alpha<0.25$; the
ferrimagnetic phase ($m=1.04$) $0.25<\alpha<0.33$; the
intermediate singlet phase $0.33<\alpha<0.43$; the ferrimagnetic
phase ($m=0.5$) $0.43<\alpha<0.6$; the singlet phase $\alpha>0.6$.
We studied the magnetization curve in all these phases and found
very specific evolution of the magnetization curve.

1) In the vicinity of the critical point ($0.25<\alpha<0.33$), the
ground state has a total spin $S_{tot}=1.04N$ and the
magnetization smoothly increases with the magnetic field from
$m=1.04$ and tends to saturation $m=3/2$ by a square-root law.

2) When $0.33<\alpha<0.43$ the ground state is singlet and the
magnetization starts to increase from zero value. At some low
value of the magnetic field $h_1$ the magnetization abruptly jumps
to $m=1/2$, then after plateau on the level $m=1/2$ at $h_2$ the
magnetization undergoes one more jump to the value slightly lower
than $m=1$. After the second jump the magnetization smoothly
increases to the saturation.

3) When $0.43<\alpha<0.6$, $h_1=0$ and the first magnetization
jump disappears, so that the magnetization starts at $m=1/2$. The
plateau on the level $m=1/2$ remains, but the magnetization jump
at $h=h_2$ transforms to the sharp square-root increase of the
magnetization curve for $h>h_2$.

4) When $\alpha>0.6$, the ground state becomes singlet and the
magnetization curve begins from $m=0$. In the region
$0.6<\alpha<2$ after short almost linear section $m\sim h$, the
magnetization undergoes a jump at some field $h_{jump}$ to a value
close to $m=1/2$. For $h>h_{jump}$ the magnetization gradually
increases to the saturation. At a certain value of $\alpha\lesssim
2$, the magnetization jump disappears, and instead the
magnetization plateau appears at the level $m=1/2$.

5) Starting from $\alpha=2$ and up to $\alpha\to\infty$, the
physical picture of the magnetization curve is as follows. At zero
magnetic field the singlet ground state of the model consists of
both basal and apical singlets. For very low magnetic fields
$h\sim\alpha^{-2}$ the apical subsystem becomes partly polarized,
while the basal subsystem remains in its singlet state. The
magnetization of the apical subsystem smoothly increases with $h$
and at some `apical saturation field' $h_{low}\approx 0.26/\alpha
^{2}$, the apical subsystem becomes fully polarized,
$m=\frac{1}{2}$. Further increase of the magnetic field
$h>h_{low}$ does not lead to the increase of the magnetization,
providing a magnetization plateau at $m=\frac{1}{2}$, until the
magnetic field overcomes the singlet-triplet gap of the basal
subsystem $h_{up}=(\Delta _{1}-\frac{1}{\alpha })$, at which the
magnetization starts to increase from the plateau at
$m=\frac{1}{2}$.

We compared the studied models with $s_a=s_b=\frac{1}{2}$ F-AF
delta chain and found that the behavior of these systems is
similar near the transition point. That is, the ground state
magnetization in the vicinity of the critical point is slightly
higher than the magnetization at the critical point,
$m_c=s_a+s_b-\frac{1}{2}$. We believe that the latter fact is
common for F-AF delta chains with any values of $s_a$ and $s_b$.

The behavior of the models far from the critical point is very
different. The ground state of the $s_a=s_b=\frac{1}{2}$ F-AF
delta chain remains ferrimagnetic for all values of $\alpha$,
while the ground state of the studied models is singlet for large
$\alpha$. This difference stems from a different type of spectrum
of basal AF chains: the gapped one for spin-$1$ and gapless for
spin-$\frac{1}{2}$. The polarization of the apical subsystem
induces a weak magnetic field acting on the basal subsystem, which
results in a weak polarization of the basal subsystem for the
$s_a=s_b=\frac{1}{2}$ F-AF delta chain, which turns out to be
energetically favorable. But for the studied model with $s_b=1$
such a weak induced magnetic field can not overcome the finite
singlet-triplet energy gap of the basal subsystem, which makes the
polarization of the apical subsystem energetically unfavorable.

\begin{acknowledgments}
The numerical calculations were carried out with use of the ALPS libraries
\cite{alps}.
\end{acknowledgments}


\begin{thebibliography}{99}
\bibitem{Diep} H.\ T.\ Diep (ed) 2013 Frustrated Spin Systems (Singapore;
World Scientific).

\bibitem{Lacrose} C.\ Lacroix, P.\ Mendels and F.\ Mila, eds., Introduction
to frustrated magnetism. Materials, Experiments, Theory(Springer-Verlag,
Berlin, 2011).

\bibitem{ModernPhysics} O.\ Derzhko, J.\ Richter, M.\ Maksymenko, Int.\ J.\
Mod.\ Phys B \textbf{29}, 153007 (2015).

\bibitem{Mac} M.\ Maksymenko, A.\ Honecker, R.\ Moessner, J.\ Richter, and
O.\ Derzhko, Phys.\ Rev.\ Lett.\ \textbf{109}, 096404 (2012).

\bibitem{Shulen} J.\ Richter, O.\ Derzhko and J.\ Schulenburg,
Phys.~Rev.~Lett. \textbf{93, }107206 (2004).

\bibitem{Zhit} M.~E.~Zhitomirsky, H.~Tsunetsugu, Phys.~Rev. B \textbf{70},
100403(R) (2004).

\bibitem{Zhit2} M.\ E.\ Zhitomirsky and H.\ Tsunetsugu, Phys.~Rev. B \textbf{%
75}, 224416 (2007).

\bibitem{Zhitomir} M.E.\ Zhitomirsky and H.\ Tsunetsugu,
Progr.~Theor.~Phys.~Suppl. \textbf{160}, 361 (2005).

\bibitem{Capponi} S.\ Capponi, O.\ Derzhko, A.\ Honecker, A.M.\ Lauchli, J.\
Richter, Phys.\ Rev.\ B \textbf{88}, 144416 (2013).

\bibitem{Balika} J.\ Richter, O.\ Krupnitska, V.\ Balika, T.\ Krokhmalski,
O.\ Derzhko, Phys.\ Rev.\ B \textbf{97}, 024405 (2018).

\bibitem{Brenig} A.\ Metavitsiadis, C.\ Psaroudaki, W.\ Brenig, Phys.\ Rev.\
B \textbf{101}, 235143 (2020).

\bibitem{malonate} C.\ Ruiz-Perez, M.\ Hernandez-Molina, P.\ Lorenzo-Luis,
F.\ Lloret, J.\ Cano, M.\ Julve M., Inorg. Chem., \textbf{39},
3845 (2000).

\bibitem{malonate1} Y.\ Inagaki, Y.\ Narumi, K.\ Kindo, H.\ Kikuchi, T.\
Kamikawa, T.\ Kunimoto, S.\ Okubo, H.\ Ohta, Y.\ Saito, M.\ Azuma,
M.\ Takano, H.\ Nojiri, M.\ Kaburagi, T.\ Tonegawa, J. Phys. Soc.
Jpn., \textbf{74}, 2831 (2005).

\bibitem{Tonegawa} T.\ Tonegawa, M.\ Kaburagi, J. Magn. Magn. Mater.,
\textbf{272--276}, 898 (2004).

\bibitem{Kaburagi} M.\ Kaburagi, T.\ Tonegawa, M.\ Kang, J. Appl. Phys.,
\textbf{97}, 10B306 (2005).

\bibitem{Ueda} R.\ Shirakami, H.\ Ueda, H.\ O.\ Jeschke, H.\ Nakano, S.\
Kobayashi, A.\ Matsuo, T.\ Sakai, N.\ Katayama, H.\ Sawa, K.\ Kindo, C.\
Michioka, K.\ Yoshimura, Phys. Rev. B \textbf{100}, 174401 (2019).

\bibitem{F10} A.\ Baniodeh, N.\ Magnani, Y.\ Lan Y., G.\ Buth, C.\ E.\
Anson, J.\ Richter, M.\ Affronte, J.\ Schnack., A.\ K.\ Powell, Npj
Quant.Mater. 2018. V. 3.1. P. 10.

\bibitem{DK} V.\ Ya.\ Krivnov, D.\ V.\ Dmitriev, S.\ Nishimoto, S.-L.\
Drechsler, and J.\ Richter, Phys. Rev. B \textbf{90}, 014441 (2014).

\bibitem{KD} D.\ V.\ Dmitriev, V.\ Ya.\ Krivnov, Phys.~Rev. B \textbf{92},
184422 (2015).

\bibitem{ferri} D.\ V.\ Dmitriev, V.\ Ya.\ Krivnov, J.\ Phys.: Condens.\
Matter \textbf{28}, 506002 (2016).

\bibitem{DKRS} D.\ V.\ Dmitriev, V.\ Ya.\ Krivnov, J.\ Richter, J.\ Schnack,
Phys.\ Rev. B \textbf{99}, 094410 (2019); Phys.\ Rev. B \textbf{101}, 054427
(2020).

\bibitem{Schnack} O.\ Derzhko, J.\ Schnack, D.\ V.\ Dmitriev, V.\ Ya.\
Krivnov, J.\ Richter, Eur.\ Phys.\ J. B \textbf{93}, 161 (2020).

\bibitem{s=1/2} T.\ Yamaguchi, S.-L.\ Drechsler, Y.\ Ohta, S.\ Nishimoto,
Phys. Rev. B \textbf{101}, 104407 (2020).

\bibitem{Rausch} R.\ Rausch, M.\ Peschke, C.\ Plorin, J.\ Schnack, C.\ Karrasch,
SciPost.\ Phys.\ \textbf{14}, 052 (2023).

\bibitem{Bert} B.\ Koteswararao, A.\ V.\ Mahajan, F.\ Bert, P.\ Mendels, J.\
Chakraborty, V.\ Singh, I.\ Dasgupta, S.\ Rayaprol, V.\ Siruguri, A.\ Hoser
and S.\ D.\ Kaushik, J.\ Phys.:Condens.\ Mat. \textbf{24}, 236001 (2012).

\bibitem{Maslova} O.\ S.\ Volkova, I.\ S.\ Maslova, R.\ Klingigiler, M.\
Abdrl-Hafiez, Y.\ C.\ Araugo, A.\ U.\ B.\ Wolter, V.\ Kataev, B.\ Buchner,
and A.\ N.\ Vasiliev, Phys.\ Rev.\ B 85, 104420 (2012).

\bibitem{Kumar} S.\ E.\ Dutton, M.\ Kumar, Z.\ G.\ Soos, C.\ L.\ Broholm,
and R.\ J.\ Cava, J.\ Phys.: Condens. Mat., 24, 166001 (2012).

\bibitem{nishimoto} T.\ Yamaguchi, Y.\ Ohta, \& S.\ Nishimoto, Phys. Rev. B,
\textbf{103}, 184410 (2021).

\bibitem{Sakai} M.\ Takahashi, T.\ Sakai, J.\ Phys.\ Soc.\ Jpn.\textbf{60, }%
760 (1991).

\bibitem{Affleck} E.\ S.\ Sorensen, I.\ Affleck, Phys. Rev. Lett \textbf{71}%
, 1633 (1993).

\bibitem{Okunishi} K.\ Okunishi, T.\ Hieida, Y.\ Akutsu, Phys. Rev. B
\textbf{59}, 6806 (1999).

\bibitem{Hodgeson} R.\ P.\ Hodgson, J.\ B.\ Parkinson, J.Phys.C\textbf{18, }%
6385 (1985).

\bibitem{Akutsu} H.\ Kiwata, Y.\ Akutsu, J.\ Phys.\ Soc.\ Jpn.\textbf{63, }%
3598 (1994).

\bibitem{White} S.\ R.\ White, Phys.\ Rev.\ Lett\textbf{.}, \textbf{69,}
2863 (1992).

\bibitem{Chandra} V.\ Ravi Chandra, D.\ Sen, N.\ B.\ Ivanov, J.\ Richter,
Phys. Rev. B \textbf{69}, 214406 (2004).

\bibitem{Grif} R.\ B.\ Griffits, Phys.\ Rev. \textbf{A133}, 768 (1964).

\bibitem{alps} F.\ Alet et al., J.\ Phys.\ Soc.\ Jpn.\ Suppl. \textbf{74},
30 (2005).
\end{thebibliography}
\end{document}